\documentclass[preprint,aps]{revtex4}
\usepackage{graphicx,epsfig}
\usepackage{natbib}

\begin{document}
\bibliographystyle{plainnat}


\title{ An essentially geometrically frustrated magnetic object: an exact solution }

\author{A.Vl. Andrianov}
\affiliation{Department of Physics, Moscow State University, Moscow 119991,
Russia }


\begin{abstract}
Starting from the archetypical geometrically frustrated magnetic objects -- equilateral triangle and tetrahedron -- 
we consider an imaginary object: a multidimensional tetrahedron with spins $1/2$ in the each vertex and equal 
Heisenberg magnetic exchange along each edge. 
Many-particle case is obtained by setting dimensionality $d$ to high numbers, hence providing
likely the most geometrically frustrated magnetic system ever possible.
This problem has an exact solution for each $d$, obtained by simple student-level approach.
As a result, this imaginary object clearly demonstrates at $d\to{}\infty$ all the features characteristic for the real geometrically frustrated magnetic systems: 
highly-degenerate ground state; 
absence of the magnetic phase transition; perfect Curie-Weiss behavior down to $T\to{}0$; and vanishingly small exchange energy 
per one spin.

\end{abstract}

\pacs{71.20.-b, 75.47.Gk, 75.50.Pp, 71.20.Nr}

\maketitle


Almost all the introductions and reviews devoted to the geometrically frustrated magnetic systems do start from the same 
sketch: an equilateral triangle with spins $1/2$ in the each vertex and antiferromagnetic exchange $J$ along each edge -- and state this system as frustrated: 
\citep{Ramirez_1994}, \citep{Schiffer_2002} etc. Another archetypical geometrically frustrated magnetic object is an equilateral tetrahedron, 
constituted by four such triangles \citep{Lee_Nature_2002}.

Starting from these objects, we consider a multidimensional equilateral tetrahedron with spins $1/2$ in the each vertex and 
obtain the many-particle case by 
setting its dimensionality $d$ to high numbers. The true thermodynamic limit corresponds to $d\to{}\infty$.
Hence, in contrast with the real systems, each spin participates in the unlimited number of frustrated triangles, 
and therefore this imaginary object might claim the title of the most
geometrically frustrated magnetic system ever.


We study an equilateral tetrahedron with spins $1/2$ in the each vertex and dimensionality $d$. It holds $d+1$ spins in total, each spin 
tied by $d$ equal edges with all the rest $d$ spins. We set Heisenberg exchange energy along each edge as $J/d$ to normalize total exchange per one spin to $J$: 
$\mathcal{H}=(J/d)\sum_{k,l}(\mathbf{s}_{k}\mathbf{s}_{l})$. Hereafter positive $J$ corresponds to the antiferromagnetic (AFM) exchange.

The non-magnetic state with the total spin number $S=0$ exists only for even number of spins, i.e. uneven $d$.
The total spin number for the given uneven $d$ varies from $S=0$ to $S_{max}=(d+1)/2$, each state $N(S,d)$-fold degenerated. 
Obviously the ground state corresponds to $S=0$  in the case of 
AFM exchange and to $S_{max}$ in the ferromagnetic FM case. Elucidation of two functions --
energy of states $\epsilon(S,d)$ and degeneration of states $N(S,d)$ -- solves the problem completely. 

The $\epsilon(S,d)$ function is easily obtained by the usual approach, see \citep{Landau_v3}: we consider the total spin 
$\mathbf{S}=\sum_{k}\mathbf{s}_{k}$, square it and and immediately obtain $S(S+1)=\sum_{k}s_{k}(s_{k}+1)+2\sum_{k,l}(\mathbf{s}_{k}\mathbf{s}_{l})=\sum_{k}(1/2\times3/2)+2\sum_{k,l}(\mathbf{s}_{k}\mathbf{s}_{l})$.
Therefore $\sum_{k,l}(\mathbf{s}_{k}\mathbf{s}_{l})=(S(S+1)-(d+1)\times3/4)/2$ and 

\begin{equation}     
\label{energy}
\epsilon(S,d)=\frac{J}{d}\sum_{k,l}(\mathbf{s}_{k}\mathbf{s}_{l})=\frac{J}{2d}\Bigl(S(S+1)-\frac{3}{4}(d+1)\Bigr)
\end{equation}
Hence $\epsilon(S_{max},d)=+J(d+1)/8$, proportional to the number of spins in the tetrahedron, and $\epsilon(0,d)=-3J(d+1)/8d$, a limited value. 
It means that in the case of AFM exchange and $d\to{}\infty$ the ground state energy per one spin, $\epsilon(0,d)/(d+1)=-3J/8d$, becomes vanishingly 
small compared to $J$. 

The degeneration function $N(S,d)$ is also easily obtained by employing the mathematical induction procedure. 
In the base case $d=0$ (i.e. single spin $1/2$) there is the only non-degenerated 
ground state $S=1/2$, so the only non-zero $N(S,d)$ value is $N(1/2,0)=1$. Next we proceed an inductive step from $d$ to $d+1$, adding therefore one spin $1/2$ to the system. As a result, each one state with $S\ne{}0$ provides two states 
with $(S+1/2)$ and $(S-1/2)$; each one state with $S=0$ provides one state with $S=1/2$. The final $N(S,d)$ values calculated this way are listed in Table.1, a modified Pascal's triangle. 

Hereafter we consider uneven $d$ values only.
Any state with the spin number $S$ has $(2S+1)$ 
space orientations in addition to the state degeneration calculated above, so for any uneven $d$ value
$\sum_{S=0}^{S_{max}}{}(2S+1)N(S,d)=2^{d+1}$, the total number of quantum states.

\begin{table}[H]
\caption{Degenerations of states for d-dimensional tetrahedron}
\begin{tabular}{c|ccccccccc}
& S= 0 & 1/2 & 1 & 3/2 & 2 & 5/2 & 3 & 7/2 &  \dots \\
\hline
d=0 & - & 1 & - & - & - & - & - & - & \dots \\
1 & 1 & - & 1 & - & - & - & - & - & \dots  \\
2 & - & 2 & - & 1 & - & - & - & - & \dots  \\
3 & 2 & - & 3 & - & 1 & - & - & - &  \dots  \\
4 & - & 5 & - & 4 & - & 1 & - & - &  \dots  \\
5 & 5 & - & 9 & - & 5 & - & 1 & - &  \dots  \\
6 & - & 14 & - & 14 & - & 6 & - & 1 &  \dots  \\
\dots & \dots & \dots & \dots & \dots & \dots & \dots & \dots & \dots & \dots   \\
15 & 1430 & - & 3432 & - & 3640 & - & 2548 & - &  \dots  \\
16 & - & 4862 & - & 7072 & - & 6188 & - & 3808 &  \dots  \\
\dots & \dots & \dots & \dots & \dots & \dots & \dots & \dots & \dots & \dots   \\
65 & $2.12\times{}10^{17}$ & - & $6.01\times{}10^{17}$ & - & $8.90\times{}10^{17}$ & - & $10.44\times{}10^{17}$ & -  & \dots  \\
66 & - & $8.13\times{}10^{17}$ & - & $14.90\times{}10^{17}$ & - & $19.34\times{}10^{17}$ & - & $21.03\times{}10^{17}$ &  \dots  \\
\dots & \dots & \dots & \dots & \dots & \dots & \dots & \dots & \dots & \dots   \\
\end{tabular}
\end{table}

On Fig.1. we present a set of $N(S)$ dependences for different $d$ values, up to 75-dimensional tetrahedron. 
The non-magnetic state with $S=0$ is always highly degenerate (while not the most degenerate state), the key feature 
of the frustrated systems. Its degeneration $N(0,d)$ rises exponentially with $d$ increase.

Now we can  write expressions for energy $E$ and susceptibility $\chi$ at the given temperature $T$ with the same ease 
as for the classical spin dimer (that corresponds to $d=1$) \citep{dimer}:

\begin{equation}     
\label{energy_T}
E(T,d)=\frac{1}{d+1}\frac{\sum_{S=0}^{S_{max}}\epsilon(S,d)(2S+1)N(S,d)\exp(-\epsilon(S,d)/kT)}{\sum_{S=0}^{S_{max}}(2S+1)N(S,d)\exp(-\epsilon(S,d)/kT)}
\end{equation}

and, assuming $\mu_{B}H\ll{}kT$,

\begin{eqnarray}
\label{suscept_T}
\chi(T,d)=\frac{1}{d+1}\frac{1}{H}\frac{\sum_{S=0}^{S_{max}}(\sum_{-S}^{S}\mu_{B}gS\exp(\mu_BgSH/kT))N(S,d)\exp(-\epsilon(S,d)/kT)}{\sum_{S=0}^{S_{max}}(2S+1)N(S,d)\exp(-\epsilon(S,d)/kT)}=\nonumber\\
\frac{\mu_{B}^2g^2}{kT(d+1)}\frac{\sum_{S=0}^{S_{max}}(\sum_{-S}^{S}S^2)N(S,d)\exp(-\epsilon(S,d)/kT)}{\sum_{S=0}^{S_{max}}(2S+1)N(S,d)\exp(-\epsilon(S,d)/kT)}=\nonumber\\
\frac{\mu_{B}^2g^2}{3kT(d+1)}\frac{\sum_{S=0}^{S_{max}}S(S+1)(2S+1)N(S,d)\exp(-\epsilon(S,d)/kT)}{\sum_{S=0}^{S_{max}}(2S+1)N(S,d)\exp(-\epsilon(S,d)/kT)}
\end{eqnarray}

Hereafter we always consider energy and susceptibility per one spin, i.e. their total values divided by $(d+1)$.
 
The high-temperature $kT\gg{}|J|$ asymptotic for (3) is always $\chi(T,d)=(\mu_{B}^2g^2/4k)/(T+J/4k)$ and does not depend on $d$. Therefore the Curie-Weiss temperature $\Theta_{CW}$ obtained in the high-temperature limit  is always equal to $(-J/4k)$ for both AFM and FM exchange, independent on the $d$ value. 
It means the proper normalization of the exchange energy we set for the model.

The $d=1$ case describes the classical spin dimer, and Eq.(3) provides the proper $\chi(T,1)=(\mu_B^2g^2/kT)/(exp(J/kT)+3)$ dependence for it \citep{dimer}.


On the Fig.2. sets of temperature dependences of energy a) obtained by Eq.2 and susceptibility c) obtained by Eq.3, both  per one spin, are demonstrated along with the temperature derivative of energy b) and inversed susceptibility d), for different dimensionality $d$ values. The trend with $d$ increase depicted by arrows. 
The main feature of Fig.4a) and b) is the clear elimination of any evidence of the magnetic phase transition with $d\to\infty$. Moreover, the exchange energy per one spin is not just small compared to $J$, but vanishes completely at $d\to\infty$. These are more pronounced frustration evidences than one can expect for the real system,
see \citep{Ramirez_2002}.

Fig.2c) and d) demonstrate trend to the perfect Curie-Weiss behavior down to $T\to{}0$ at $d\to\infty$, well below $-\Theta_{CW}=J/4k$ (that corresponds to $0.25$ on the $x$ scale) -- 
that is also a key feature for the frustrated magnetic systems. 


In addition we consider the non-frustrated case of FM exchange, $J<0$.
Seemingly this problem at $d\to\infty$ is perfect for the mean-field approach, because the coordination number, equal to $d$, is unlimitedly high. This approach, obviously, predicts magnetic ordering at the Curie-Weiss temperature $\Theta_{CW}=T_{C}=|J|/4k$; $E=0$ and exact Curie-Weiss behavior above it; drop in $dE/dkT$ equal to $3/2$ and divergence in $\chi(T)$ at $T_C$ -- all the features familiar for the mean-field approach to an ordinary ferromagnetic transition \citep{Stanley_MFT}. Nevertheless the results obtained by our calculations, while clearly tend to this mean-field prediction, are still far from it even at relatively high $d$ values.

On the Fig.3. sets of temperature dependences of energy a) obtained by Eq.2 and susceptibility c) obtained by Eq.3, both  per one spin, are demonstrated along with temperature derivative of energy b) and inversed susceptibility d) for different dimensionality $d$ values. We see that even a 75-dimensional tetrahedron, with its 76 spins, behaves far enough from the mean-field prediction, while trend to this behavior with $d$ increase, depicted by arrows, is clear. Likely it is caused by the limited size of our object that is always equal to one interatomic distance, so correlation length has no room to diverge.



In conclusion, we consider an imaginary geometrically frustrated magnetic system that has a simple and transparent exact solution.
It demonstrates, in a pronounced way, all the features characteristic for the real frustrated systems: 
highly-frustrated non-magnetic ground state; 
absence of the magnetic phase transition down to the lowest temperatures;
vanishingly small exchange energy per one spin;
exact Curie-Weiss behavior down to the lowest temperatures.

We hope this model might provide a useful intuition to the behavior of the geometrically  frustrated magnetic systems.


Author is indebted to M.Tsypin and M.Katsnelson for useful discussions.


\newpage

\begin{figure} [tbh]  
 \resizebox{1.1\columnwidth}{!}{\includegraphics*{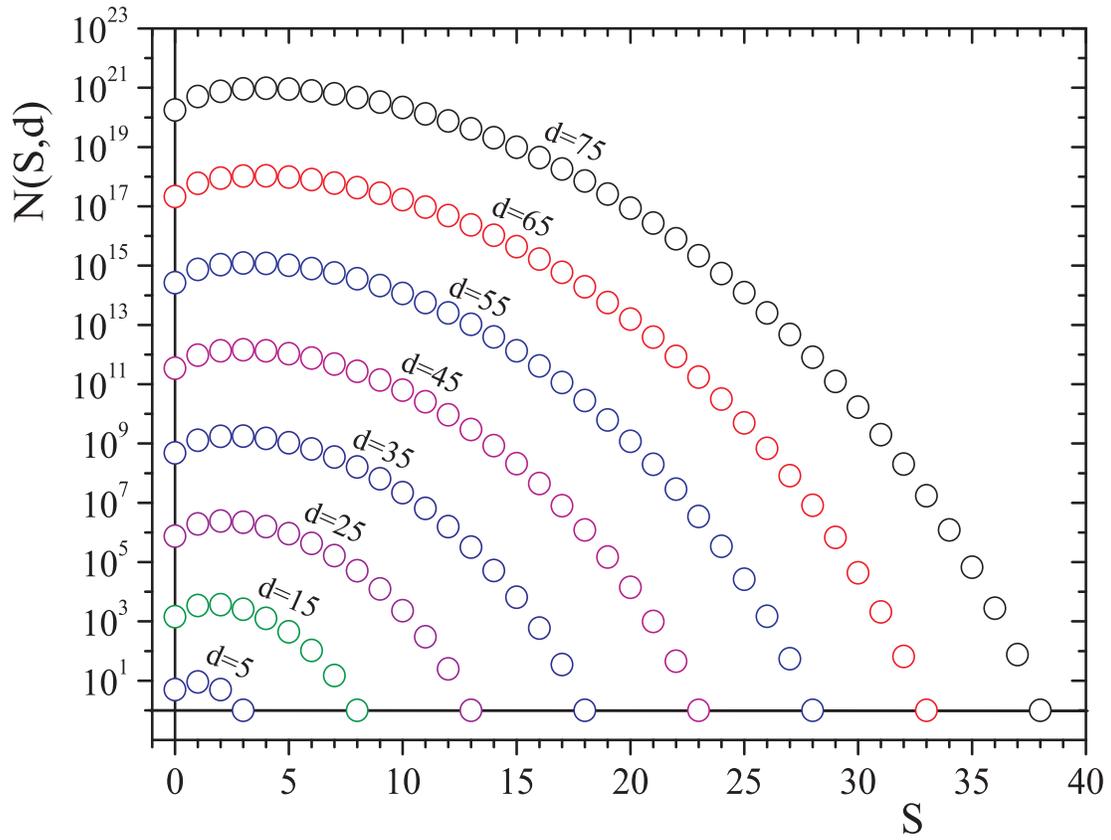}}
\caption{
(Color online.)
The degeneration of states $N(S,d)$ for different $d$ values, labeled over the dependences (see text).
}
\end{figure}

\begin{figure} [tbh]   
\resizebox{1.1\columnwidth}{!}{\includegraphics*{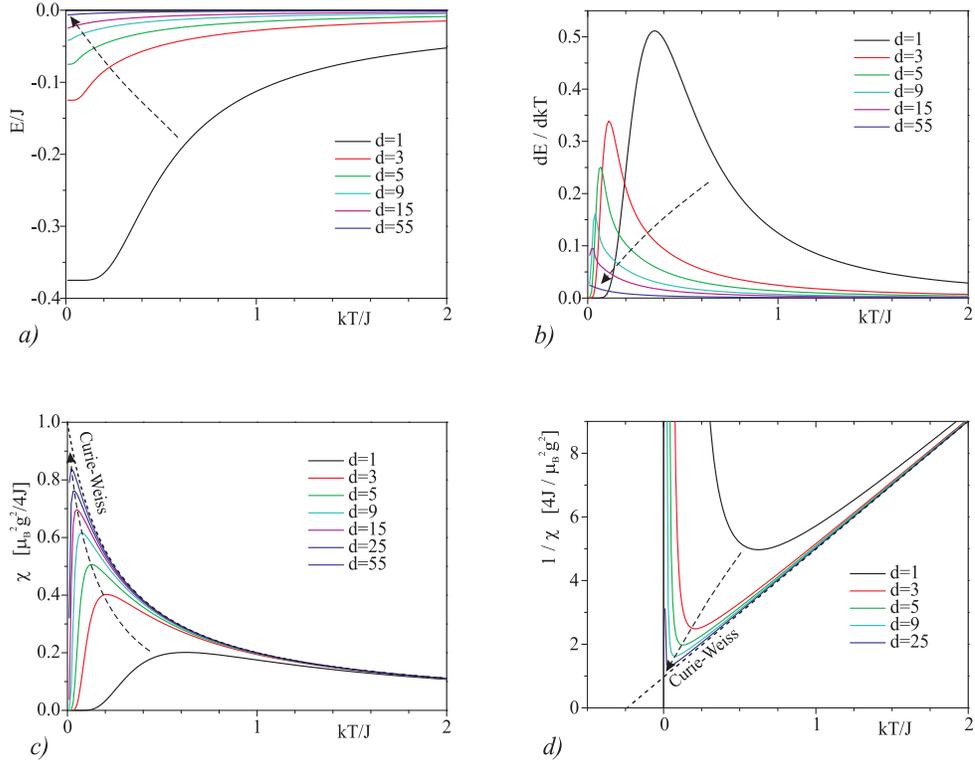}}
\caption{
(Color online.) Antiferromagnetic exchange $J>0$, Curie-Weiss temperature is $-J/4k$. Sets of temperature dependences for different $d$: 
a) energy per one spin; b) its temperature derivative; c) susceptibility per one spin; 
d) inversed susceptibility  per one spin. Short-dashed line is a Curie-Weiss dependence.
The trend with $d$ increase depicted by arrows. 
}
\end{figure}

\begin{figure}   
\resizebox{1.1\columnwidth}{!}{\includegraphics*{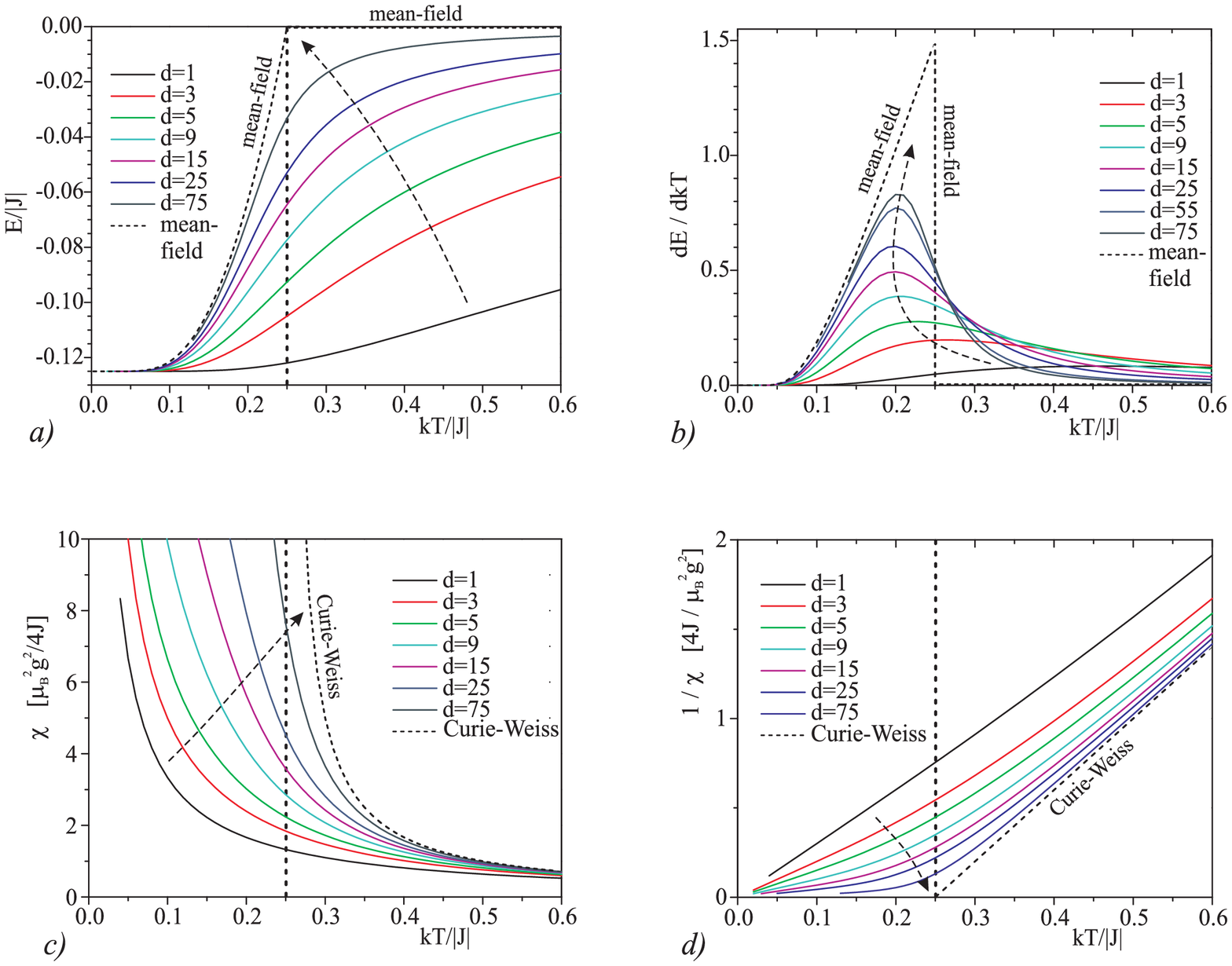}}
\caption{
(Color online.) Ferromagnetic exchange $J<0$, Curie-Weiss temperature is $|J|/4k$, marked by vertical dotted line. 
Sets of temperature dependences for different $d$: 
a) energy per one spin; b) its temperature derivative; c) susceptibility per one spin; 
d) inversed susceptibility  per one spin. Short-dashed line is a mean-field approximation dependence,  
a Curie-Weiss dependence for c), d) also.
The trend with $d$ increase depicted by arrows. 
}
\end{figure}


\begin{references}


\bibitem{Ramirez_1994}   
A.P.Ramirez,
Annu. Rev. Mater. Sci. {\bf 24} 453 (1994)


\bibitem{Schiffer_2002}   
P. Schiffer,
Nature {\bf 420} 35 (2002)

\bibitem{Lee_Nature_2002} 
S.-H. Lee, C. Broholm, W. Ratcliff, G. Gasparovic, Q. Huang, T. H. Kim, S.-W. Cheong,
Nature {\bf 418}, 856 (2002)




\bibitem{Landau_v3} 
L. D. Landau, E. M. Lifshitz,
{\it Quantum Mechanics: Non-Relativistic Theory}, 101,
(Pergamon Press Ltd., 1977)

\bibitem{dimer} 
I. S. Jacobs, J. W. Bray, H. R. Hart, Jr., L. V. Interrante, J. S. Kasper, G. D. Watkins,
D. E. Prober, J. C. Bonner,
Phys. Rev. B {\bf 14} 3036 (1976)

\bibitem{Ramirez_2002}  
A.P. Ramirez,
{\it Handbook of Magnetic Materials}, {\bf 13}, 423, 
ed. by K.H. J. Buschow,
(Elsevier Science B.V. Amsterdam : North-Holland, 2001)

\bibitem{Stanley_MFT} 
H.E. Stanley,
{\it Intoduction to phase transitions and critical phenomena}, 86,
(Oxford Univ. Press, 1987)


\end{references}
\end{document}